\documentclass[journal,twoside,web]{ieeecolor}
\usepackage{tmi}
\usepackage{cite}
\usepackage{amsmath,amssymb,amsfonts}
\usepackage{graphicx}
\usepackage{textcomp}
\usepackage{booktabs}
\usepackage[noend]{algpseudocode}
\usepackage{algorithmicx,algorithm}
\usepackage{bm}
\usepackage{bbding}
\def\BibTeX{{\rm B\kern-.05em{\sc i\kern-.025em b}\kern-.08em
		T\kern-.1667em\lower.7ex\hbox{E}\kern-.125emX}}
\begin{document}
	\title{
		%[1] 3D Vessel Segmentation under the Guidance of 2D Structure-agnostic Vessel Annotations
		%[2] 3D vessel segmentation under little supervision: guided by annotations of other 2D vessels
		3D Vessel Segmentation with Limited Guidance of 2D Structure-agnostic Vessel Annotations
	}
	\author{Huai Chen, Xiuying Wang, Lisheng Wang
		\thanks{Huai Chen and Lisheng Wang (corresponding author) are with Institute of Image Processing and Pattern Recognition, Department of Automation, Shanghai Jiao Tong University, Shanghai 200240, P. R. China. E-mail: lswang@sjtu.edu.cn}
		\thanks{Xiuying Wang is with the School of Computer Science, The University of Sydney, Sydney, NSW 2006, Australia.}
	}
	
	\maketitle
	
	\begin{abstract}
		
		Delineating 3D blood vessels is essential for clinical diagnosis and treatment, however, is challenging due to complex structure variations and varied imaging conditions. Supervised deep learning has demonstrated its superior capacity in automatic 3D vessel segmentation. However, the reliance on expensive 3D manual annotations and limited capacity for annotation reuse hinder the clinical applications of supervised models. To avoid the repetitive and laborious annotating process and make full use of existing vascular annotations, this paper proposes a novel 3D shape-guided local discrimination (3D-SLD) model for 3D vascular segmentation under limited guidance from public 2D vessel annotations. The primary hypothesis is that 3D vessels are composed of semantically similar voxels and often exhibit tree-shaped morphology. Accordingly, the 3D region discrimination loss is firstly proposed to learn the discriminative representation measuring voxel-wise similarities and cluster semantically consistent voxels to form the candidate 3D vascular segmentation in unlabeled images; secondly, based on the similarity of the tree-shaped morphology between 2D and 3D vessels, the Crop-and-Overlap strategy is presented to generate reference masks from the existing 2D structure-agnostic vessel annotations, which are fit for varied vascular structures, and the adversarial loss is introduced to guide the 3D vessels with the tree-shaped morphology; and thirdly, the temporal consistency loss is proposed to foster the training stability and keep the model updated smoothly. To further enhance the model's robustness and reliability, the orientation-invariant CNN module and Reliability-Refinement algorithm are presented. Experimental results from the public 3D cerebrovascular and 3D arterial tree datasets demonstrate that our model achieves comparable effectiveness against nine deep learning supervised models.
	\end{abstract}
	
	
	\section{Introduction}
	\label{sec:introduction}
	
	\IEEEPARstart{B}{lood} vessels are the major component of the circulatory system, whose functions are to deliver nutrients and oxygen through the body and transport the waste from tissues/organs \cite{chaurasia2004human}. Blood vessels are distributed in almost all anatomical structures, and the changes in their geometry and morphology are closely associated with various vascular-related diseases \cite{devasagayam2016cerebral,pu2022automated,he2020learning,ciecholewski2021computational}. Take the cerebral vessel as an example, it is one of the most important vascular structures and its morphological abnormality is a sign of cerebrovascular diseases, including cerebral thrombosis and cerebral hemangioma \cite{devasagayam2016cerebral,flemming2017population}. Furthermore, the visualization of cerebrovascular networks plays an important role in optimal treatment and neurosurgery planning \cite{xia20223d}. Therefore, vessel analysis and visualization, especially for 3D vessels, is critically important for clinical diagnosis and treatment. For 3D vessel analysis and visualization, vascular segmentation in 3D medical images, such as time-of-flight magnetic resonance angiography (TOF-MRA) images \cite{chen2022attention,chen2022generative}, is the prerequisite. However, due to the complex morphology variations and varied imaging modalities and protocols, manual segmentation for 3D vessels is time-consuming and laborious \cite{lesage2009review,kozinski2020tracing}. Moreover, the reliance on annotators' experience and other human factors restrict the repeatability and reproducibility of the manual delineation \cite{moccia2018blood}. To tackle these issues, automatic 3D vessel segmentation methods are in high demand \cite{chen2022generative,chen2022attention,xia20223d}.
	
	Due to their superior performance, deep learning-based methods have recently dominated the field of automatic 3D vessel segmentation \cite{moccia2018blood,guo2021cerebrovascular,zhang2020cerebrovascular,li2022robust}. However, as the mainstream deep learning-based methods, the supervised deep-learning models rely on large amounts of 3D manual annotations, which hinders their clinical application and slows down the model building \cite{xia20223d,li20213d,chen2022attention}. To address this issue, Subramaniam et al. propose a model to generate 3D TOF-MRA volumes accompanied by vascular annotations based on generative adversarial networks (GAN), which can enlarge the training samples based on limited annotated data \cite{subramaniam2022generating}; Chen et al. show a semi-supervised framework to alleviate the demand for annotated data \cite{chen2022generative}; and kozinski et al. \cite{kozinski2020tracing} present a model to realize 3D vessel segmentation only based on annotations on the associated 2D Maximum Intensity Projections (MIPs), accelerating the annotating process. However, a certain amount of labeled 3D data is still needed to initialize the model in \cite{subramaniam2022generating,chen2022generative}; and vessel masks in the associated 2D MIPs are still required in \cite{kozinski2020tracing}. Even worse, all of the supervised deep-learning models and \cite{subramaniam2022generating,chen2022generative,kozinski2020tracing} lack the ability of annotation reuse, which means the annotated data only support the segmentation model for images of the same vascular structure and medical modality. Consequently, the burdensome annotating work is unavoidable when building 3D vascular segmentation models for images of new advanced medical modalities or new structures.
	
	\begin{figure}[]
		%\centering	
		%\includegraphics[width=\textwidth]{figures/main_stream.png}
		\includegraphics[width=0.97\columnwidth]{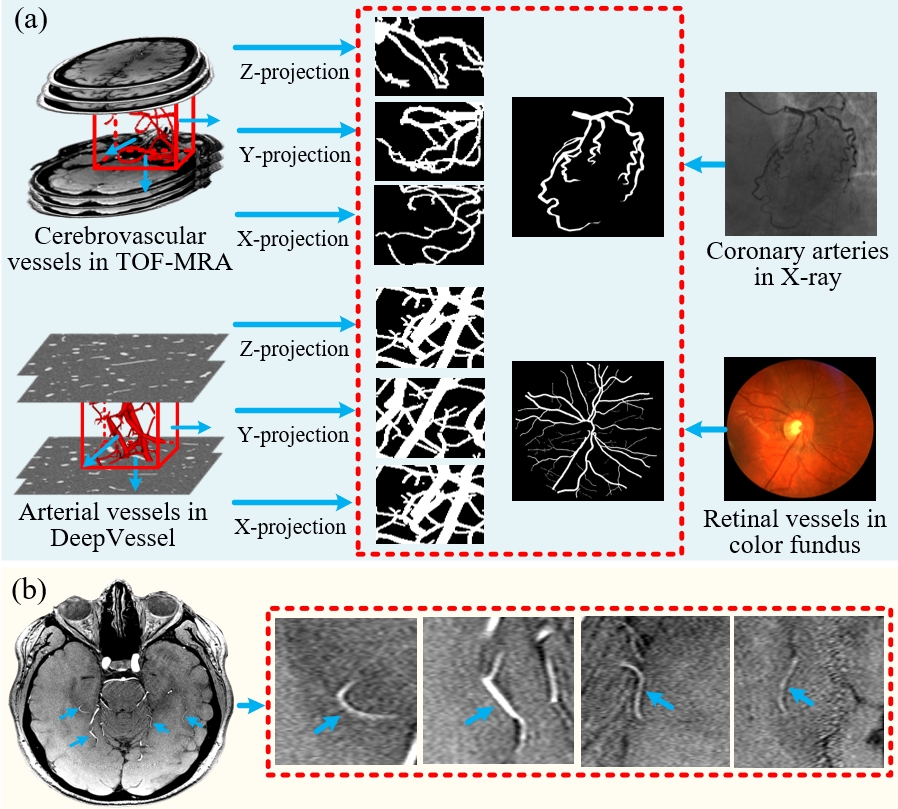}
		\caption{An illustration to show the similar tree-shaped morphology between the MIPs of 3D and 2D vessel masks (shown in (a)), and the semantic consistency among vessel voxels (shown in (b)).} 
		\label{fig:illustration_for_tree_shape_and_semantic_consistency}
	\end{figure}
	%DeepVessel: a synthetic 3D vessel dataset \cite{tetteh2020deepvesselnet}.
	
	To avoid the expensive and repetitive 3D vascular annotating process, we propose the 3D shape-guided local discrimination model (3D-SLD) for 3D vascular segmentation, which has the capacity of reusing the morphology prior from available 2D structure-agnostic vessel annotations. Its primary hypothesis is that 3D vessels can be segmented by harnessing the morphology consistency between 2D and 3D vessels with the semantic consistency of vessel voxels. Specifically, vessel masks in 2D medical images share similar tree-shaped morphologies with the 2D MIPs of 3D vessel masks \cite{li2022human} (as illustrated in Fig.~\ref{fig:illustration_for_tree_shape_and_semantic_consistency}.(a)). Therefore, it is cost-effective to reuse the morphology knowledge from existing structure-agnostic 2D vessel masks, such as 2D coronary arteries (CA) annotations \cite{ma2021self} and 2D retinal vessels (RV) annotations \cite{fraz2012ensemble}, as the reference to constrain the 2D MIPs of the 3D vessel. This constraint on 2D MIPs can then in turn further retain the tree-shaped morphology of the whole 3D vessels \cite{kozinski2020tracing}. Secondly, as shown in Fig.~\ref{fig:illustration_for_tree_shape_and_semantic_consistency}.(b), vascular voxels share similar semantic patterns, such as intensity and texture \cite{tortora2018principles}. Based on this property, we can cluster semantically consistent voxels to form the candidate segmentation for 3D vessels. 
	
	Based on the hypothesis, there are mainly three challenges to be addressed when constructing 3D-SLD. Firstly, discriminative representation is urgently needed to measure the semantic similarity between voxels; secondly, due to the variation in density and thickness of different vascular structures, it is challenging to effectively utilize the morphology prior of 2D vessels to guide the segmentation for different 3D vascular structures; thirdly, the mechanism of training models without annotations specifically designed for target vascular structures makes the training lack the supervision information, which easily leads to training unstability. To address these issues, the 3D region discrimination loss is firstly presented to learn a discriminative space based on unlabeled 3D images, by which, the voxel-wise semantic similarities can be measured and semantically consistent voxels can be clustered to form the candidate 3D vascular region. Secondly, the strategy of Crop-and-Overlap is presented to generate reference masks from existing 2D vessel masks, which are fit for different vascular structures, and the adversarial loss is further introduced to align the distribution of 3D segmentation’s MIPs with these references. By doing this, the whole 3D segmentation is constrained to own the tree-shaped morphology. Thirdly, the temporal consistency loss is proposed to keep the model consistent with the corresponding exponential moving average (EMA) model, which keeps the model updated stably. Through the above implementations, 3D vessels, which are tree-shaped structures with internal consistency, can be stably identified. Moreover, considering the orientation-invariant property of 3D vessels, we propose the orientation-invariant convolutional module (ori-CNN) to learn features robust to visual orientation changes for better 3D vessel representation; based on the hypothesis that predictions with high confidence are reliable, Reliability-Refinement algorithm is presented to improve the model's reliability by using reliable pseudo labels. 
	
	Our method has been validated in the important and challenging tasks of 3D cerebrovascular segmentation \cite{chen2022attention} and 3D arterial tree segmentation \cite{tetteh2020deepvesselnet}. Only guided by 2D CA/RV masks, our method obtains results comparable to nine deep learning supervised models, showing its effectiveness and superior annotation reuse capacity. 
	
	In summary, the major contributions are three folds:
	
	\begin{itemize}
		\item[(a)] %We propose a novel 3D shape-guided local discrimination model based on the morphology similarity between 2D and 3D vessels, which effectively segments 3D vessels by only reusing the morphology prior of 2D structure-agnostic vessel annotations.
		We propose a novel 3D shape-guided local discrimination model to effectively segment 3D vessels based on the morphology prior of 2D structure-agnostic vessels and unlabeled 3D images. It shows essential clinical values to avoid the expensive and repetitive manual annotating and make full use of the existing vascular annotations.
		
		\item[(b)] To robustly identify the 3D vessels, the novel 3D region discrimination loss, adversarial loss and temporal consistency loss are respectively proposed to cluster semantically consistent voxels, ensure predictions’ tree-shaped morphologies and enhance the training stability.
		\item[(c)] To further enhance the model’s reliability and robustness to visual orientation changes, Reliability-Refinement algorithm and the ori-CNN module are presented.
	\end{itemize}
	
	\section{Related work}
	\label{sec:related work}
	\subsection{3D vessel segmentation}
	Recent 3D vessel segmentation algorithms can be generally categorized into traditional models and deep learning-based models. By analyzing the prior knowledge of intensity \cite{frangi1998multiscale,foruzan2012hessian} and shape \cite{lee2015adaptive,liang20153d}, most of the traditional models extract hand-craft and customized features to describe and delineate the 3D vessels. These methods are usually annotation-free and possess good interpretability. However, the hand-craft features are insufficiently robust, which results in the limited performance of traditional methods \cite{moccia2018blood}. To capture more robust and more comprehensive features, deep learning-based models have been proposed and become the hot research topic for 3D vessel segmentation \cite{li20213d,kozinski2020tracing,chen2022generative,xia20223d}. To realize cerebrovascular segmentation in TOF-MRA, Chen et al. \cite{chen2022attention} propose A-SegAN, which consists of a A-SegS for segmentation and a A-SegC for discriminating predictions from the ground truth. Li et al. \cite{li20213d} focus on improving the connectivity of vessels and present a graph attention network (GAT) to model the graphical connectivity information. However, most of the recent deep learning-based methods are greedy for expensive manual annotations. 
	
	Although some recent deep learning-based work \cite{subramaniam2022generating,chen2022generative,kozinski2020tracing} has proposed models to alleviate the demand for annotations or accelerate the delineating process, they still need experts to provide annotations for target vascular structure. Even worse, all of them lack the ability of annotation reuse, which leads to repetitive annotating when building models for different vascular structures. Comparatively, our 3D-SLD is cost-efficient and can realize 3D vessel segmentation based on raw 3D images and existing 2D vessel masks. Furthermore, it owns the superior annotation reuse ability so that one 2D vessel mask dataset owns the capacity of guiding the 3D vessel segmentation for images of different modalities and organs.
	
	\subsection{Cross-domain knowledge transfer}
	Cross-domain knowledge transfer aims to enhance models' performance in target domains based on the shared knowledge from highly relevant domains. In medical image processing, it can be mainly divided into domain adaptation \cite{wu2021unsupervised,guan2021domain} to alleviate the degradation caused by varied imaging conditions and synthetic segmentation \cite{huo2018synseg,chen2020anatomy} to adapt the trained model to test images focusing on the same body part but with different medical modalities.
	
	%It can be mainly divided into domain adaptation \cite{wu2021unsupervised,guan2021domain} and synthetic segmentation \cite{huo2018synseg,chen2020anatomy} in medical image processing. Respectively, domain adaptation attempts to alleviate the degradation caused by varied imaging conditions; and synthetic segmentation aims to adapt the trained model to test images focusing on the same body part but with different medical modalities.
	
	Besides the aforementioned methods for transferring knowledge between images of the same modality or images of the same body part, our previous work \cite{chen2021unsupervised2,chen2021unsupervised} proposes models for transferring morphology knowledge between structures with similar shapes to realize target segmentation in 2D medical images. This work follows the idea of identifying the target by the guidance of morphology prior from other structures. In contrast, this work focuses on the essential task of 3D vessel segmentation and makes a breakthrough to realize 3D vessel segmentation by reusing the morphology prior from 2D structure-agnostic vessel annotations. Furthermore, the temporal consistency loss, ori-CNN and Reliability-Refinement algorithm are respectively presented to enhance the training stability, robustness and reliability of the model.
	
	%\vspace{1cm}
	\section{Methodology}
	\begin{figure*}[t]
		\centering	
		\includegraphics[width=0.96\textwidth]{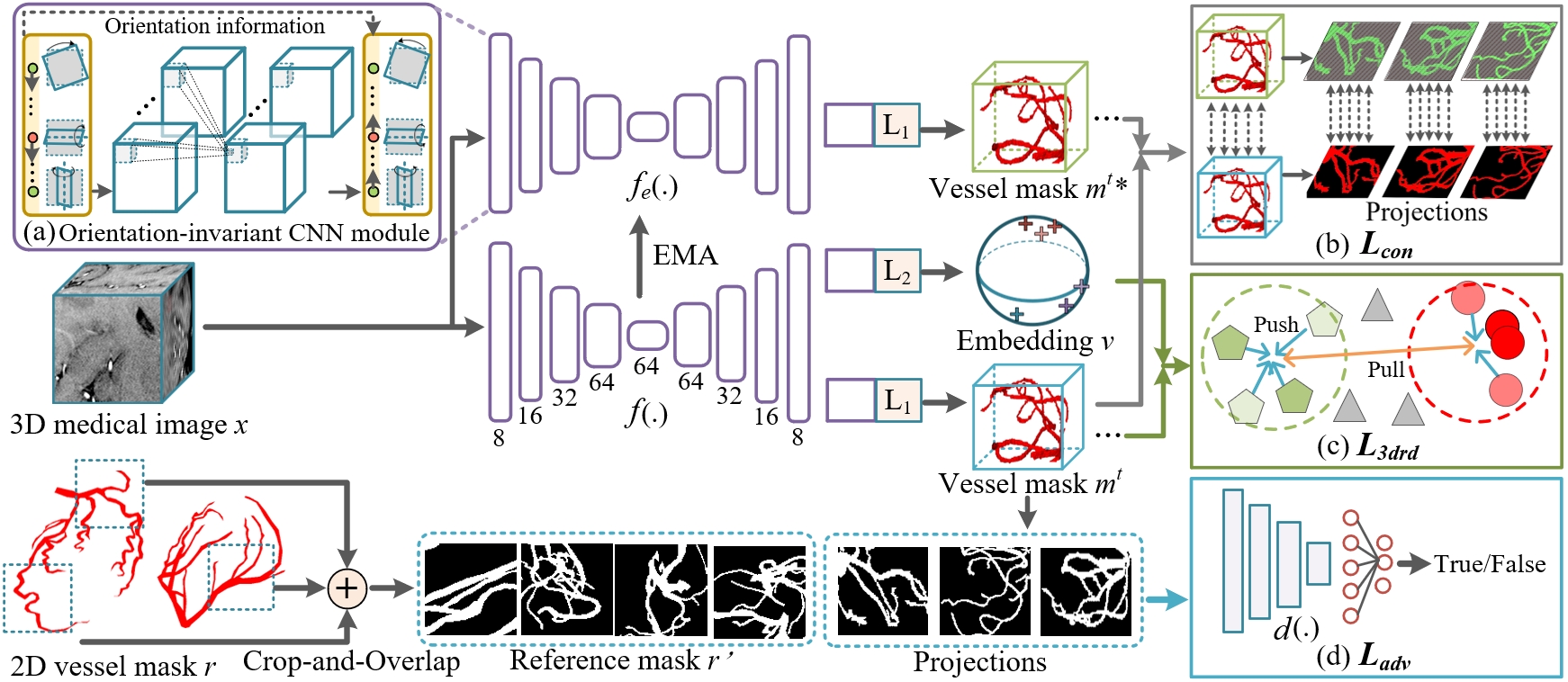}
		\caption{An illustration of 3D-SLD. The backbone $f(\cdot)$ is a 3D U-Net (the skip connections are omitted for simplicity) and its basic component is ori-CNN (shown in (a)). $f_e(\cdot)$ is the EMA model, reflecting the temporal information of $f(\cdot)$. $d(\cdot)$ is the discriminator. (b), (c) and (d) respectively show $L_{con}$ to ensure the temporal consistency, $L_{3drd}$ to cluster semantically consistent voxels and $L_{adv}$ to guide the tree-shaped segmentation.} 
		\label{fig:framework}
	\end{figure*}
	
	The overall framework of 3D-SLD is illustrated in Fig.~\ref{fig:framework}. The input is 3D medical volumes $X=\{x_1,x_2,...,x_{N_1}\}$ and 2D vessel masks $R=\{r_1,r_2,...,r_{N_2}\}$. We respectively set the depth, height and width of the medical volume as $D$, $H$ and $W$, i.e., $x_n\in \mathbb{R}^ {D\times H\times W}$. Similarly, the 2D mask is denoted as $r_n \in {\mathbb{R}^ {H\times W}}$. When feeding $X$ into the model, the backbone $f(\cdot)$ can generate segmentation masks and voxel-wise embedding, by fusing which we can get the 3D region discrimination loss $L_{3drd}$ to optimize $f(\cdot)$ to generate candidate 3D vessel segmentation clustering semantically similar voxels (shown in Fig.~\ref{fig:framework}.(c) and articulated in Sec.~\ref{section:3D_rd}). Simultaneously, 2D vessel masks are utilized to generate reference masks by Crop-and-Overlap, based on which the adversarial loss $L_{adv}$ is presented to guide the tree-liked morphology of 3D vessel masks (shown in Fig.~\ref{fig:framework}.(d) and articulated in Sec.~\ref{section:shape_guided}). On the other hand, the EMA model $f_e(\cdot)$ is further constructed to reflect the temporal information of $f(\cdot)$ and is utilized to calculate temporal consistency loss $L_{con}$ to ensure the training stability of $f(\cdot)$ (shown in Fig.~\ref{fig:framework}.(b) and described in Sec.~\ref{section:EMA_model}). By optimizing $f(\cdot)$ to minimize these losses, 3D vessels, which are tree-like structures composed of semantically similar voxels, can be stably identified. It is worth noting that the details of Reliability-Refinement algorithm are introduced in Sec.~\ref{section:Reliable_Refinement} to further enhance the model's reliability.

	%In the following, the detailed definition and notation of the backbone, including its basic component of ori-CNN, are articulated in Sec. III-A. In Sec. III-B, we present the 3D region discrimination loss to learn local discriminative features and cluster semantically consistent voxels. The elucidation of utilizing 2D vessel masks to ensure the tree-like morphology of the segmentation mask is shown in Sec. III-C. Then, the temporal consistency loss is elucidated in Sec. III-D. Finally, the Reliable-Refinement algorithm is introduced in Sec. III-E
	
	\subsection{The Architecture of the Backbone}
	\label{section:method_backbone}
	
	The backbone of our model $f(\cdot)$ is designed as a 3D U-Net style network, which can effectively fuse low-level and high-level features and is the recommended architecture for medical segmentation models \cite{ronneberger2015u}. Then, a clustering module and an embedding module are set to realize voxel-wise embedding and clustering. When feeding an image $x_n$ into $f(\cdot)$, we can get the embedding $v_n$ and the segmentation mask $m_n$. The detailed formulation is as follows:
	\begin{equation}
		\label{eqution:backbone}
		\begin{aligned}
			v_n,m_n&=f(x_n),
		\end{aligned}	
	\end{equation}
	where $v_n(d,h,w) \in \mathbb{R}^{K}$ represents the $K$-dimensional embedded vector for voxel $x_n(d,h,w)$; $m_n^c(d,h,w) (0<c \leq C)$ denotes the probability of categorizing voxel $x_n(d,h,w)$ into $c$-th semantic class and there are a total of $C$ classes. It is worth noting that $v_n(d,h,w)$ is normalized by an $l_2$-normalization layer so that the embedding is projected on a hypersphere space and $m_n(d,h,w)$ is normalized by an $l_1$-normalization layer so that the sum of the classification probability is 1, i.e., $\left\|v_n(d,h,w)\right\|_2=1$ and $\left\|m_n(d,h,w)\right\|_1=1$.
	
	\noindent \textbf{Orientation-invariant CNN Module:} The 3D vessel owns the orientation-invariant property, which means the identity of 3D vessels should not be affected by the change in the visual orientation. Therefore, we design the ori-CNN module to learn orientation invariant features to better describe 3D vessels.
	
	Specifically, as shown in Fig.~\ref{fig:framework}.(a), in the ori-CNN module, features from the previous module are firstly randomly augmented by multiple orientation-related augmentations to change the orientation property. Then, the augmented features are further processed by CNN layers to get deeper features. Finally, the output features are restored to their original orientation according to the augmentation settings. By setting ori-CNN as the basic block for $f(\cdot)$, the CNN layer is fit for handling the orientation-agnostic input, and the generated features are robust to orientation changes.
	
	%In this paper, each ori-CNN contains two CNN layers and the augmentation of flipping and $90^\circ$/$180^\circ$/$270^\circ$-rotation in the X/Y/Z-axis are utilized.   
	
	\subsection{$\mathcal{L}_{3drd}$ to generate internally consistent regions}
	\label{section:3D_rd}
	3D region discrimination loss $\mathcal{L}_{3drd}$ is designed to optimize $f(\cdot)$ to ensure the generated $m_n$ clusters semantically similar voxels. It is based on contrastive learning \cite{ye2019unsupervised,chen2021unsupervised} and its theoretical basis is that medical images of the same modality and the same body part commonly share similar structures or tissues \cite{chen2021unsupervised2}. Therefore, each of these images can be segmented into multiple internally consistent regions and similar regions can be found in each image. Based on this, we firstly fuse the semantic embedding $v_n$ and the segmentation mask $m_n$ to represent the semantic information for each clustered region. Then, the 3D region discrimination loss is designed to optimize $f(\cdot)$ to correctly assign the semantically consistent regions into the same class, by which, the discriminative embedding space can be learnt to project semantically similar voxels together and the segmentation clusters similar voxels. 
	
	\noindent \textbf{The embedded vector for clustered regions and semantic classes:} According to the description in Sec.~\ref{section:method_backbone}, a high value of $m^c_n(d,h,w)$ indicates that the associated voxel $x_n(d,h,w)$ has high confidence of belonging to the $c$-th semantic class. Therefore, the embedding of this voxel reasonably plays a more important role when presenting the $c$-th semantic region in $x_n$. Following this analysis, we denote the embedded vector of the $c$-th semantic region in $x_n$ as $t_{nc}$, and define it as the weighted sum of voxel embedding as:
	\begin{equation}
		\label{equation:tnc}
		t_{nc}=\frac{\sum_{d,h,w}m^c_{n}(d,h,w)v_n(d,h,w)}{\left\|\sum_{d,h,w}m^c_{n}(d,h,w)v_n(d,h,w)\right\|_2}.
	\end{equation}
	After getting the embedding for each region in each image, we can further denote the prototype (cluster centroid) of $c$-th semantic class as $T_c$, which is defined as the sum of $t_{nc} (0 < n \leq N_1)$. The detailed formulation is as follows:
	\begin{equation}
		T_{c}=\frac{\sum_{n}t_{nc}}{\left\|\sum_{n}t_{nc}\right\|_2}.
	\end{equation}
	
	\noindent \textbf{The definition of $\mathcal{L}_{3drd}$:} Then, we can obtain the probability of correctly assigning $t_{nc}$ to $T_c$ as follows:
	\begin{equation}
		\label{equation:Pnc}
		P_{nc}=\frac{e^{sim(t_{nc},T_c)/\tau}}{\sum_{0\leq c' <C}e^{sim(t_{nc},T_{c'})/\tau}},
	\end{equation}
	where $sim(a,b)$ measures the similarity between vectors, and it is set as the cosine similarity, i.e., $sim(a,b)=\frac{a^\top b}{\left\|a\right\|_2 \left\|b\right\|_2}$; $\tau$ is set as 0.1 to control the concentration level of the sample distribution \cite{hinton2015distilling}. Since $t_{nc}$ and $T_c$ are both on the hypersphere space, $sim(a,b)$ can be simplified as $sim(a,b)=a^\top b$. 
	
	After that, we can obtain the joint probability of making correct assignments as $\prod \limits_{n,c} P_{nc}$, and its negative log-likelihood is defined as the region discrimination loss $\mathcal{L}_{3drd}$, i.e.:
	\begin{equation}
		\label{equation:rd}
		\mathcal{L}_{3drd}=-\sum_{n,c}log(P_{nc}).
	\end{equation}
	
	\noindent \textbf{Analyses:} Eq.~\ref{equation:Pnc} can be rewritten as:
	\begin{equation}
		\label{equation:rewritting}
		P_{nc}=\frac{1}{1+\sum_{c'\neq c} \frac{e^{sim(t_{nc},T_{c'})/\tau}}{e^{sim(t_{nc},T_c)/\tau}}}.
	\end{equation}  
	By minimizing $\mathcal{L}_{3drd}$ (Eq.~\ref{equation:rd}), $P_{nc}$ is driven to go up. According to Eq.~\ref{equation:rewritting}, $sim(t_{nc},T_{c'}) (c'\neq c)$ is forced to be smaller and $sim(t_{nc},T_{c})$ attends to become bigger, i.e., $t_{nc}$ is pushed away from $T_{c'}$ and is pulled toward to $T_{c}$. To satisfy this optimization target, voxels with embedding $v_n(d,h,w)$ being highly similar to $T_n$ are optimized to have high $m^c_n(d,h,w)$ values (according to the definition of $t_{nc}$ in Eq.~\ref{equation:tnc}). That is, semantically consistent voxels can be clustered together into the same semantic class (prototype).
	
	\subsection{$\mathcal{L}_{adv}$ to ensure the tree-like morphology of segmentation}
	\label{section:shape_guided}
	We assign the $t$-th segmentation mask $m^t_n$ as the segmentation mask for 3D vessels ($t$ is a scalar parameter and $0<t \leq C$). $\mathcal{L}_{3drd}$ described in Sec.~\ref{section:3D_rd} ensure $m^t_n$ clusters semantically consistent voxels. This section aims to further add the morphology constraint from 2D vessel masks to $m^t_n$, so that $m^t_n$ owns the tree-like shape and can identify 3D vessels. 
	
	\noindent \textbf{Generating MIPs of $x^t_n$:} For $m^t_n$, we respectively denote its MIPs along the X, Y and Z axes as $s^1_n$, $s^2_n$ and $s^3_n$, i.e.:
	\begin{equation}
		\begin{aligned}
			s^1_n(d,h)&=\max_{w}(m^t_n(d,h,w)),\\
			s^2_n(d,w)&=\max_{h}(m^t_n(d,h,w)),\\
			s^3_n(h,w)&=\max_{d}(m^t_n(d,h,w)).\\
		\end{aligned}
	\end{equation}
	We set the generated MIPs set as $S=\{s^1_1,s^2_1,s^3_1,s^1_2,s^2_2,...\}$.
	
	\noindent \textbf{Crop-and-Overlap to generate reference masks:} Although MIPs of 3D vessel masks and 2D vessel masks share similar tree-shaped morphologies, they still differ in density and thickness (shown by the mask visualization in Fig.~\ref{fig:illustration_for_tree_shape_and_semantic_consistency}.(a)). To address this issue, we propose a simple but effective strategy, named Crop-and-Overlap, to generate reference masks fitting different 3D vascular structures. As shown in the bottom part of Fig.~\ref{fig:framework}, when feeding 2D vessel masks into 3D-SLD, multiple patches are randomly cropped and then overlapped to construct the reference mask $r'$. By implementing this generating strategy and adjusting the crop scale of patches and the overlap number, we can obtain reference masks $R'=\{r'_1,r'_2,...\}$ with varied vascular thicknesses and densities.
	
	\noindent \textbf{Adversarial training for the shape distribution alignment:} Adversarial training has been proven to minimize the Jensen-Shannon divergence between two distributions and can be utilized to align distributions \cite{goodfellow2020generative}. Therefore, we apply adversarial training to align the shape distribution of MIPs of 3D vessel masks with the shape distribution of 2D vessel masks, further ensuring that 3D vessel masks own a tree-like shape.
	
	Concretely, we firstly set a discriminator $d(\cdot)$. Then, $S$ and $R'$ are fed into $d(\cdot)$, and the classification loss $\mathcal{L}_d$ is introduced to optimize $d(\cdot)$ such that samples from $R'$ and $S$ are correctly distinguished. The detailed definition of $\mathcal{L}_d$ is as follows:
	\begin{equation}
		\begin{aligned}
			\label{eq:Ld}
			\mathcal{L}_d&=\mathcal{L}_{bce}(d(R'),1)+\mathcal{L}_{bce}(d(S),0),\\
			\mathcal{L}_{bce}(Y,Y')&=-\frac{1}{N}\sum\limits_{n}(y_n\text{log}y'_n+(1-y_n)\text{log}(1-y'_n)).
		\end{aligned}
	\end{equation}
	Meanwhile, $f(\cdot)$ is optimized by minimizing the adversarial loss $\mathcal{L}_{adv}$ to cheat $d(\cdot)$. The definition of $\mathcal{L}_{adv}$ is as follows:
	\begin{equation}
		\label{eq:Ladv}
		\mathcal{L}_{adv}=\mathcal{L}_{bce}(d(S),1).
	\end{equation} 
	By simultaneously optimizing $d(\cdot)$ and $f(\cdot)$, the MIPs of $x^t_n$ is motivated to own a similar tree-like morphology as reference masks. Since the corresponding MIPs can provide sufficient constraints for whole 3D vessels \cite{kozinski2020tracing}, $x^t_n$ is encouraged to own the expected 3D tree-like shape.
	
	\subsection{$\mathcal{L}_{con}$ to enhance the model's training stability}
	\label{section:EMA_model}
	By simultaneously minimizing $\mathcal{L}_{3drd}$ and $\mathcal{L}_{adv}$ described in the previous two sections, we can optimize $f(\cdot)$ to identify internal semantically consistent regions with a tree-shaped morphology in the 3D medical volume, i.e., obtain 3D vessel masks. However, the absence of manual annotations and the mechanism of only utilizing 2D vessel masks as the guidance for 3D vessels make the training of $f(\cdot)$ lack supervised information. As a result, $f(\cdot)$ may fluctuate sharply during training and its predictions rapidly change (the detailed analyses can be seen in Sec.~\ref{sec:ablation_study}). To address this issue, the temporal consistency loss is proposed.  
	
	\noindent \textbf{Constructing the EMA model:} Concretely, the EMA model can reflect the model's temporal information and is demonstrated to obtain better predictions \cite{tarvainen2017mean}; therefore, the EMA model of $f(\cdot)$, denoted as $f_e(\cdot)$, is constructed to provide temporal information to keep $f(\cdot)$ smoothly updated. In training step $i$, we denote the weights of $f(\cdot)$ as $\theta_i$ and the weights of $f_e(\cdot)$ as $\theta ^*_i$. Then, $\theta ^*_i$ is updated as follows:
	\begin{equation}
		\theta ^*_i=\alpha_i \theta ^*_{i-1} +(1-\alpha_i) \theta_{i},
	\end{equation}
	where $\alpha_i$ is the smoothing coefficient to control the update rate. Since the reliability of $f_e(.)$ is improved with the increase of training steps, we design $\alpha_i$ as a ramp-up weight as follows:
	\begin{equation}
		\alpha_i=1-\frac{1}{i/I+1},
	\end{equation}
	where $I$ is the iteration step in one epoch.
	
	\noindent \textbf{The definition of temporal consistency loss $\mathcal{L}_{con}$:} The EMA model $f_e(\cdot)$ can reflect the historical information of $f(\cdot)$, and we expect the output of $f(\cdot)$ is not severely different from the previous version to ensure temporal stability. Therefore, we leverage the output of $f_e(\cdot)$ and its corresponding MIPs as the guidance for $f(\cdot)$. The definition of $L_{con}$ is as follows:
	\begin{equation}
		\label{eq:Lcon}
		\begin{aligned}
			&\mathcal{L}_{con}=\mathcal{L}_{dice}(M^t,M^{*t})+\mathcal{L}_{dice}(M^t_{MIPs},M^{*t}_{MIPs}),\\
			&\ \ \ \ \ \ \ \ \ \ \ \ \ \ \ M^t=\{m^{t}_1,m^{t}_2,...,m^{t}_n,...\},\\		
			&\ \ \ \ \ \ \ \ \ \ \ \ \ M^{*t}=\{m^{*t}_{1},m^{*t}_{2},...,m^{*t}_{n},...\},\\
			&\ \ \ \ \ \ \ \ \  M^t_{MIPs}=\{s^1_1,s^2_1,s^3_1,...,s^1_n,s^2_n,s^3_n,...\},\\
			&\ \ \ \ \ \ \ M^{*t}_{MIPs}=\{s^{*1}_1,s^{*2}_1,s^{*3}_1,...,s^{*1}_n,s^{*2}_n,s^{*3}_n,...\},\\
			&\ \ \ \ \ \ \ \ \  \mathcal{L}_{dice}(Y,Y')=\frac{-2\sum_{n}y'_ny_n+\epsilon}{\sum_{n}y'_n+\sum_{n}y_n+\epsilon},\\
		\end{aligned}
	\end{equation}
	where $m^t_n$ is the 3D vessel segmentation mask generated by $f(\cdot)$, and $s^1_n,s^2_n,s^3_n$ are the corresponding MIPs along the X, Y and Z axes (as described in Sec.~\ref{section:shape_guided}). Similarly, $m^{*t}_n$ is the 3D vessel segmentation mask generated by $f_e(\cdot)$, and the corresponding MIPs are $s^{*1}_n,s^{*2}_n,s^{*3}_n$. $\mathcal{L}_{dice}$ measures the overlap rate, and the smoothness term $\epsilon$ is set as 1. After implementing this loss, $f(\cdot)$ is motivated to keep consistent with its EMA model $f_e(\cdot)$, which ensures its smooth update.

	\subsection{Reliability-Refinement algorithm to improve the model}
	\label{section:Reliable_Refinement}
	\begin{figure}[b]
		\centering	
		\includegraphics[width=0.8\columnwidth]{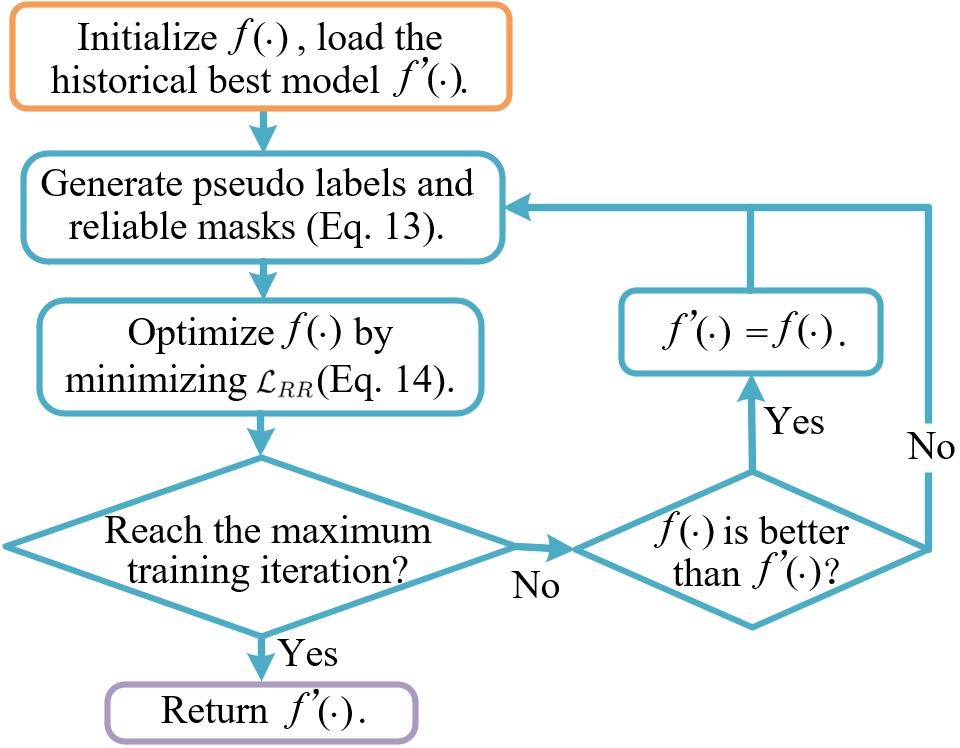}
		\caption{The flow diagram of Reliability-Refinement algorithm.} 
		\label{fig:reliable_retraining}
	\end{figure}
	
	Previous parts can successfully construct a stable and robust 3D vessel segmentation model based on the shape guidance from 2D vessel masks. In this part, we take a deeper step to improve the model's reliability. The main idea is to identify the high-confidence predictions as pseudo labels and use them to further refine the model. The flow diagram of the algorithm is shown in Fig.~\ref{fig:reliable_retraining}.
	
	\noindent \textbf{Generating pseudo labels and confident masks:} Specifically, during training, the historical best model $f'(\cdot)$ is updated in real-time. Then, the pseudo label for $x_n$ is generated by integrating the predictions of $f'(\cdot)$ for multiple $A(x_n)$ ($A(x_n)$ denotes the augmented version of $x_n$). Since the integrated result with a high probability or a low probability (close to 1 or 0) of being classified as the vessel region is more confident, we identify these regions as reliable masks. We denote the pseudo label set as $Y_p=\{y_1,y_2,...,y_n,...\}$ and the reliable mask set as $Q=\{q_1,q_2,...,q_n,...\}$. Their definitions are:
	\begin{equation}
		\begin{aligned}
			&\ \ \ \ \ \ y'_n=mean\{\underbrace{f'(A_1(x_n)),f'(A_2(x_n)),...}_{N_e\ samples}\},\\
			&\ \ \ \ \ \ \ \  y_n(d,h,w)=
			\begin{cases}
				0,y'_n(d,h,w)\leq 0.5,\\
				1,y'_n(d,h,w)>0.5,
			\end{cases}\\
			& \ \ \ \ q_n(d,h,w)=
			\begin{cases}
				0,\ (1-v_c)<y'_n(d,h,w)<v_c,\\
				1,\ otherwise,\\
			\end{cases}\\
		\end{aligned}
	\end{equation}
	where $v_c$ is the confidence rate used to filter out insufficiently reliable predictions and $N_e$ is the number of augmented data.
	
	\noindent \textbf{The definition of the Reliability-Refinement loss $\mathcal{L}_{RR}$:} $\mathcal{L}_{RR}$ is based on the dice loss, and the reliable mask is utilized to identify the confident pseudo labels. Its definition is as follows:
	\begin{equation}
		\label{eq:LRR}
		\begin{aligned}
			Y=&\{f(x_1),f(x_2),...,f(x_n),...\},\\
			\mathcal{L}_{RR}(Y,Y_p,Q)&=\frac{-2\sum_{n}y_nf(x_n)q_n+\epsilon}{\sum_{n}y_nq_n+\sum_{n}f(x_n)q_n+\epsilon},\\
		\end{aligned}
	\end{equation}
	
	\section{Experimental Results and Analyses}
	\label{sec:experimental results}
	\subsection{Datasets and preprocessing} 
	
	Two 3D vessel datasets, containing the cerebrovascular segmentation dataset \cite{chen2022attention} and the DeepVessel dataset \cite{tetteh2020deepvesselnet}, are utilized to evaluate our method; and public 2D vessel dataset of XCAD \cite{ma2021self} is leveraged to generate 2D reference masks. 
	
	\noindent $\bullet$ \textbf{Cerebrovascular segmentation dataset} contains 45 TOF-MRA volumes with cerebrovascular annotations. We re-sample the spatial resolution to $0.4mm \times 0.4mm \times 0.4mm$, and the final spatial size is $674 \times 674 \times 184$. We further provide brain masks to filter out no-brain parts. Five and one annotated instances are split as the test data and validation data. And the remaining raw images are set as the training data.
	
	\noindent $\bullet$ \textbf{The DeepVessel dataset} is a 3D arterial tree dataset for vessel segmentation, centerline prediction and bifurcation detection. There are 136 annotated images in total and the spatial size is $325 \times 304 \times 600$. Twenty-six and one annotated instances are split as the test data and validation data. And the remaining raw images are set as the training data.
	
	\noindent $\bullet$ \textbf{XCAD} is an X-ray angiography coronary artery disease dataset, containing 126 public images with coronary artery annotations. We resize annotation masks to $512\times 512$ and use them as the morphology reference for our model. 
	
	\subsection{Implementation details}
	\subsubsection{Network architectures} The backbone $f(\cdot)$ is made up of five encoder blocks and four decoder blocks, and each block is an ori-CNN module with two $3\times 3\times 3$ CNN layers. Each of the first four encoder blocks is followed by a 3D $2\times 2\times 2$ max-pooling layer to enlarge the receptive field. And before each decoder block, a 3D $2\times 2\times 2$ up-sampling layer is set to restore the resolution. Both the embedding module and the clustering module are composed of two CNN layers. The channel number of the embedding module and the clustering module are respectively 16 and 8, i.e., $K=16$ and $C=8$. $f_e(\cdot)$ shares the same architecture as $f(\cdot)$. The discriminator $d(\cdot)$ is a simple classifier composed of four 2D $3\times 3$ CNN layers and two fully connected layers. The first two CNN layers are followed by a $2\times 2$ max-pooling layer and the final CNN layer is followed by a global averaging layer. The channel numbers for each layer are $16,32,32,32,32,1$, and the final out is activated by the Sigmoid.
	
	\subsubsection{Training process} The training flow can be mainly divided into three stages as follows:
	
	\noindent $\bullet$ \textbf{Pre-training $f(\cdot)$ by 3D patch discrimination:} Similar to \cite{chen2021unsupervised2,chen2021unsupervised}, 3D version patch discrimination with hypersphere mixup is utilized to pre-train the model to own the initial discriminative ability. Its main idea is to learn patch-wise discriminative features by forcing $f(\cdot)$ to distinguish patch instances. Please refer to \cite{chen2021unsupervised} for more details on the patch discrimination loss and hypersphere mixup loss. Specifically, we divide an image volume into $4\times 4\times 4$ patches and optimize $f(\cdot)$ by minimizing the 3D version losses, which are denoted as $\mathcal{L}_{3dPd}$ and $\mathcal{L}_{3dHm}$. The training loss is defined as:
	\begin{equation}
		\mathcal{L}=\mathcal{L}_{3dPd}+\mathcal{L}_{3dHm}.
	\end{equation}
	
	\noindent $\bullet$ \textbf{Training 3D-SLD to obtain initial segmentation:} The model pre-trained by patch discrimination is utilized to initialize our 3D-SLD. When training the 3D-SLD, $d(\cdot)$ is optimized by minimizing $\mathcal{L}_{d}$ (Eq.~\ref{eq:Ld}), and the total loss to optimize $f(\cdot)$ is defined as follows:
	\begin{equation}
		\mathcal{L}=\mathcal{L}_{3dPd}+\mathcal{L}_{3dHm}+10\mathcal{L}_{3drd}+\mathcal{L}_{adv}+\lambda_i \mathcal{L}_{con}+0.1\mathcal{L}_{e},
	\end{equation}
	where $\mathcal{L}_{3drd}$, $\mathcal{L}_{adv}$ and $\mathcal{L}_{con}$ are respectively articulated by Eq.~\ref{equation:rd}, Eq.~\ref{eq:Ladv} and Eq.~\ref{eq:Lcon}. The goal of $\mathcal{L}_{e}$ is to minimize the entropy of $m_n$ in order to motivate the segmentation predictions to own high confidence. Its definition is:
	\begin{equation}
		\begin{aligned}
			\mathcal{L}_{e}=&-[\frac{1}{NCDHW}\sum_{n,c,d,h,w}m^c_{n}(d,h,w)\text{log}(m^c_{n}(d,h,w))+\\
			&\sum_{n,c,d,h,w}(1-m^c_{n}(d,h,w))\text{log}(1-m^c_{n}(d,h,w))].
		\end{aligned}
	\end{equation}
	Since $f_e(\cdot)$ generates more reliable predictions as the increase of training step $i$, the coefficient $\lambda_i$ for $\mathcal{L}_{con}$ is designed as a ramp-up weight. Its detailed definition is as follows:
	\begin{equation}
		\lambda_i=min(4i/I,10),
	\end{equation}
	where $I$ is the total number of training steps in one epoch. 
	
	\noindent $\bullet$ \textbf{Refinement on the model's reliability:} Reliability-Refinement algorithm is used to further enhance the model's reliability (described in Sec.~\ref{section:Reliable_Refinement}). To construct the pseudo label, the predictions of 8 different augmented versions of one medical volume are generated ($N_e=8$). And $v_c$ is set as 0.9 to identify the reliable mask. $\mathcal{L}_{RR}$ (Eq.~\ref{eq:LRR}) is used to optimize $f(\cdot)$ with the supervision of pseudo labels.
	
	\subsubsection{Experimental settings} The training epochs for 3 training stages are respectively set as 100, 40 and 60, and models are optimized 200 iterations (i.e., $I=200$) in one epoch. In each iteration, 3 batches of data are used as the training data, and each batch contains four cropped images with the size of $96\times 96\times 96$ and two mixup images. Meanwhile, 24 2D reference masks are generated by Crop-and-Overlap (described in Sec.~\ref{section:shape_guided}) and are resized to $96\times 96$. The optimizer is set as the Adam with the learning rate of 0.001. 
	
	Our experiments are conducted on a workstation platform equipped with 4$\times$ NVIDIA GeForce GTX 1080 GPU. The operating system is Ubuntu 18.04LTS and the codes are implemented with PyTorch 1.10.1.
	
	\subsubsection{Quantitative metrics:} Dice Similarity Coefficient (DSC), precision rate (PR) and recall rate (RR) are utilized to evaluate the experimental results. Their definitions are as follows:
	\begin{equation}
		\begin{aligned}
			DSC=\frac{2|Y\cap Y'|}{|Y|+|Y'|}, PR=\frac{TP}{TP+FP},RR=\frac{TP}{TP+FN},\\
		\end{aligned}
	\end{equation}
	where $Y$ is the binary prediction and $Y'$ is the ground truth; $TP$ is the number of true positives; $FP$ is the number of false positives; and $FN$ is the number of false negatives.
	
	\subsection{Experimental Comparisons with Supervised Deep Learning Models}
	\label{sec:experimental-comparisons}
	We evaluate the performance of our proposed segmentation model through the experimental comparisons with 9 supervised deep learning segmentation models including 3 general segmentation models and 6 models designated for 3D vessel segmentation: (1) \textbf{General segmentation models}: \textit{DSN} \cite{dou20173d}, \textit{3D-UNet} \cite{cciccek20163d} and \textit{V-Net} \cite{milletari2016v} are popular and commonly used 3D deep learning segmentation models. (2) \textbf{3D vessel segmentation models}: \textit{DeepVesselNet-Fcn}, \textit{DeepVesselNet-3DUNet} and \textit{DeepVesselNet-VNet} \cite{tetteh2020deepvesselnet} propose the cross-hair filter, which enhances vascular segmentation models' performance and alleviates the time and GPU consumption. \textit{3DUNet-MIPs-1}, \textit{3DUNet-MIPs-2} and \textit{3DUNet-MIPs-3} \cite{kozinski2020tracing} are cost-efficiency methods, which only need associated 2D annotations in MIPs to train the model. Respectively, 3DUNet-MIPs-1, 3DUNet-MIPs-2 and 3DUNet-MIPs-3 are based on annotations on 1 MIP, 2 MIPs and 3 MIPs.
	
	\begin{figure*}[t]
		\centering	
		\includegraphics[width=0.98\textwidth]{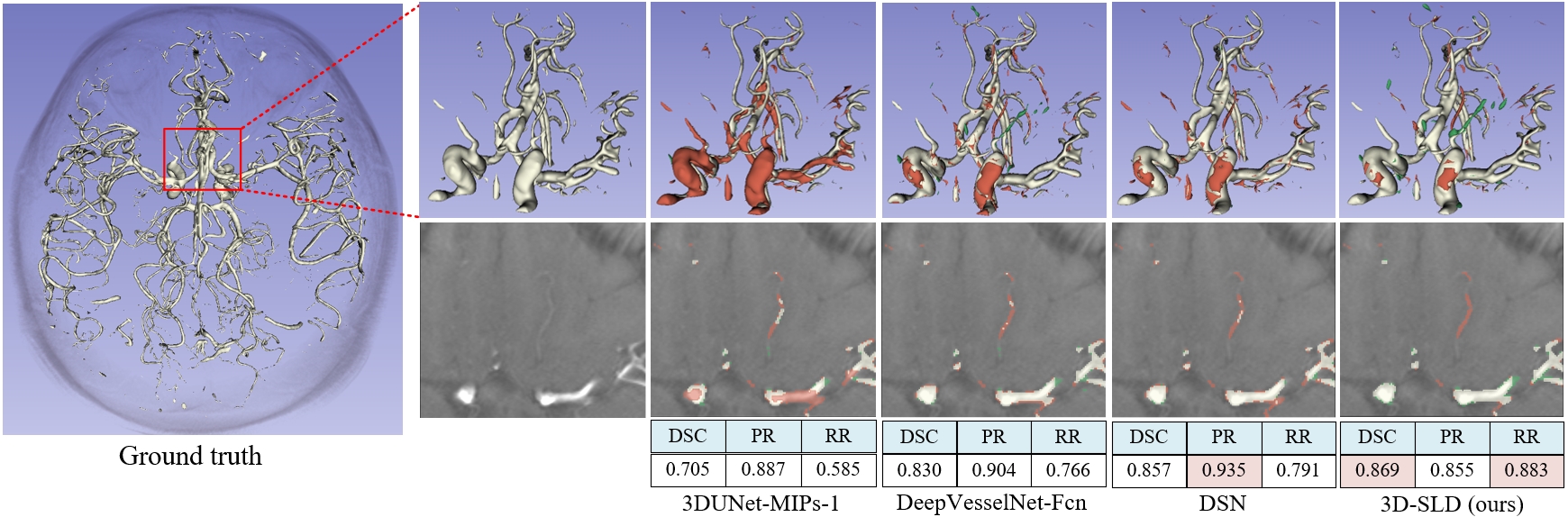}
		\caption{The visualization of cerebrovascular segmentation results. The first row and the second row respectively show the prediction in 3D volumes and 2D slices. Regions covered with \textcolor{red}{red} and \textcolor{green}{green} respectively show false negatives and false positives.} 
		\label{fig:comparative_visualization}
	\end{figure*}

	\begin{table}[]%Comparison of results of downstream segmenting tasks
		\caption{Comparison of the cerebrovascular segmentation results and the arterial tree segmentation results.}
		\centering
		\renewcommand\tabcolsep{3.5pt}
		\label{table:brain_vessel_segmentation_results}
		\begin{tabular}{l|ccc}
			\toprule[1.3pt]
			Methods & DSC(\%)  & PR(\%) & RR(\%) \\
			\hline
			
			\multicolumn{4}{c}{Part 1: the cerebrovascular segmentation results}\\
			\hline
			\hline
			
			&\multicolumn{3}{|c}{General supervised models}\\
			\hline
			DSN & $79.12$ & $91.45$ & $69.91$ \\
			3D-UNet & $82.8$ & $90.46$ & $76.54$ \\
			V-Net & $83.46$ & $88.44$ & $79.2$ \\
			\hline
			&\multicolumn{3}{|c}{3D vessel segmentation models}\\
			\hline
			DeepVesselNet-Fcn & $73.89$ & $80.45$ & $68.98$ \\
			DeepVesselNet-3DUNet & $81.45$ & $91.87$ & $73.33$ \\
			DeepVesselNet-VNet & $83.53$ & $89.4$ & $78.52$ \\
			3DUNet-MIPs-1 & $77.35$ & $86.81$ & $69.91$ \\
			3DUNet-MIPs-2 &$81.96$& $88.25$ & $76.67$ \\
			3DUNet-MIPs-3 &$83.21$& $89.0$ & $78.29$ \\
			
			\hline
			3D-SLD (ours)& $79.04$ & $86.38$ & $73.05$ \\
			\hline
			\hline
			
			\multicolumn{4}{c}{Part 2: the arterial tree segmentation results}\\
			\hline
			\hline
			&\multicolumn{3}{|c}{General supervised models}\\
			\hline
			DSN & $97.98$ & $98.47$ & $97.49$ \\
			3D-UNet & $99.96$ & $99.94$ & $99.97$ \\
			V-Net & $99.96$ & $99.95$ & $99.98$ \\
			\hline
			&\multicolumn{3}{|c}{3D vessel segmentation models}\\
			\hline
			DeepVesselNet-Fcn & $99.48$ & $99.45$ & $99.5$ \\
			DeepVesselNet-3DUNet & $99.93$ & $99.9$ & $99.96$ \\
			DeepVesselNet-VNet & $99.96$ & $99.94$ & $99.97$ \\
			3DUNet-MIPs-1 & $94.64$ & $99.85$ & $89.94$ \\
			3DUNet-MIPs-2 & $99.04$ & $99.97$ & $98.12$ \\
			3DUNet-MIPs-3 & $99.59$ & $99.97$ & $99.21$ \\
			\hline
			3D-SLD (ours)& {$98.6$} & {$98.63$} & {$98.57$} \\
			\hline
			\hline
		\end{tabular}
	\end{table}
	
	From the comparative results shown in Table.~\ref{table:brain_vessel_segmentation_results} and the visualization of cerebrovascular segmentation results shown in Fig.~\ref{fig:comparative_visualization}, we have the following observations:
	
	1) Firstly, our method, trained only under the guidance of 2D CA masks, obtains results comparable to the supervised methods. As shown in Table.~\ref{table:brain_vessel_segmentation_results}, in cerebrovascular segmentation, our 3D-SLD outperforms DeepVeselNet-Fcn and 3DUNet-MIPs-1 in DSC by 5.15\% and 1.69\%, and gets similar results to DSN (DSN owns the better PR and 3D-SLD owns the better RR). As shown in Fig.~\ref{fig:comparative_visualization}, DSN and our method are superior to DeepVeselNet-Fcn and 3DUNet-MIPs-1, and respectively work better in alleviating false positives and false negatives. In arterial tree segmentation, 3D-SLD outperforms DSN and 3DUNet-MIPs-1. Compared with these expensive supervised methods, which are trained with 39 annotated TOF-MRA images or 109 annotated DeepVessel images, our method is cost-efficient and only based on public 2D coronary artery annotations. Thus, the comparable results sufficiently show the effectiveness of our model. 
	
	2) Secondly, our method shows a superior capacity for annotation reuse. To realize segmentation for different 3D vascular structures, supervised methods need to construct different annotated datasets for each structure. For example, annotated TOF-MRA images are needed for supervised models to realize cerebrovascular segmentation in TOF-MRA, while, annotated DeepVessel images are needed for supervised models to realize arterial tree segmentation in DeepVessel images. Comparatively, in our 3D-SLD, the public 2D CA annotations own the power to guide the segmentation for different vascular structures, including 3D cerebral vessels and 3D arterial trees.

	\begin{table}[]%Comparison of results of downstream segmenting tasks
		\caption{Comparative experiments of models based on different reference datasets and one reference mask.}
		\centering
		\renewcommand\tabcolsep{3.5pt}
		\label{table:different_reference}
		\begin{tabular}{l|ccc}
			\toprule[1.3pt]
			Methods & DSC(\%)  & PR(\%) & RR(\%) \\
			\hline
			\hline
			\multicolumn{4}{c}{Part 1: models based on full reference masks}\\
			\hline
			&\multicolumn{3}{c}{The cerebrovascular segmentation}\\
			\hline
			3D-SLD (RV-guided) & $77.57$ & $81.28$ & $\bm{74.59}$ \\
			3D-SLD (CA-guided)& $\bm{79.04}$ & $\bm{86.38}$ & $73.05$ \\
			\hline
			
			&\multicolumn{3}{c}{The arterial tree segmentation}\\
			\hline
			3D-SLD (RV-guided) & $98.56$ & \bm{$98.64}$ & $98.47$\\
			3D-SLD (CA-guided)& \bm{$98.6$} & $98.63$ & \bm{$98.57$}  \\
			\hline
			\hline
			
			\multicolumn{4}{c}{Part 2: models based on one reference mask}\\
			\hline
			&\multicolumn{3}{c}{The cerebrovascular segmentation}\\
			\hline
			3D-SLD (RV-guided) & $73.46$ & $78.96$ & \bm{$69.04$}\\
			3D-SLD (CA-guided)& \bm{$74.99$} &\bm{$84.56$}& $67.56$ \\
			\hline
			
			&\multicolumn{3}{c}{The arterial tree segmentation}\\
			\hline
			3D-SLD (RV-guided) & $96.80$ &$95.91$ & \bm{$97.71$} \\
			3D-SLD (CA-guided)& \bm{$97.43$} & \bm{$97.40$} & $97.46$\\
			\hline
			\hline
			
		\end{tabular}
	\end{table}
	
	\subsection{Analyses on models with different reference datasets} 
	\label{sec:different-reference}
	To analyze the model's sensibility to the selection of reference masks, besides the 2D CA annotations in XCAD, we also utilize 28 retinal vessel (RV) annotations in the color fundus dataset of CHASEDB \cite{fraz2012ensemble} as the reference masks (visualizations of 2D CA and RV masks can be seen in Fig.~\ref{sec:introduction}). The comparison results of CA-guided models and RV-guided models are shown in part 1 of Table.~\ref{table:different_reference}. We can see that although CA and RV masks are varied in the vascular density and thickness, CA-guided and RV-guided models can obtain similar results; especially, the arterial tree segmentation results of the RV-guided model and the CA-guided model are extremely close. This observation shows that our model is robust to the selection of 2D reference vascular masks.
	
	\subsection{Analyses on models with extremely limited references}
	\label{sec:limited-references}
	We analyze the performance of our model with extremely limited references, i.e., with only one 2D reference mask. The experimental results are shown in part 2 of Table.~\ref{table:different_reference}. From the results, we can see that our models still work in this severe condition. Especially, in the arterial segmentation task, the result gap between models trained with full reference masks and models trained with only one reference mask is small. This observation shows the robustness of our model and demonstrates that our model is capable of fully utilizing the prior knowledge provided by reference masks. 
	
	\subsection{Analyses on ablation experiments}
	\label{sec:ablation_study}
	\begin{table}[]%Ablation experiments of the MRA segmentation
		\caption{Ablation experiments of the cerebrovascular segmentation and the arterial tree segmentation.}
		\centering
		\renewcommand\tabcolsep{3.5pt}
		\label{table:Ablation_brain_vessel}
		\begin{tabular}{l|ccc|ccc}
			\toprule[1.3pt]
			Model & ori-CNN & $\mathcal{L}_{con}$ & RR & DSC(\%)  & PR(\%) & RR(\%)\\
			\hline
			\hline
			&\multicolumn{6}{c}{The cerebrovascular segmentation}\\
			\hline	
			A&\XSolidBrush &\XSolidBrush &\XSolidBrush &$ 56.25 $&$ 60.45 $&$ 53.25$ \\
			B&\Checkmark &\XSolidBrush &\XSolidBrush &$ 61.19 $&$ 65.29 $&$ 57.96$ \\
			C&\XSolidBrush &\Checkmark &\XSolidBrush &$ 70.39 $&$ 80.98 $&$ 62.53$ \\
			D&\Checkmark &\Checkmark &\XSolidBrush & $76.53$ & $83.38$ & $70.99$ \\
			E&\Checkmark &\Checkmark &\Checkmark & \bm{$79.04$} & \bm{$86.38$} & \bm{$73.05$} \\
			\hline
			\hline
			
			&\multicolumn{6}{c}{The arterial tree segmentation}\\
			\hline
			A&\XSolidBrush &\XSolidBrush &\XSolidBrush & $84.44$ & $82.1$ & $86.92$ \\
			B&\Checkmark &\XSolidBrush &\XSolidBrush & $94.61$ & $91.0$ & $98.51$ \\
			C&\XSolidBrush &\Checkmark &\XSolidBrush & $95.44$ & $98.38$ & $92.67$ \\
			D&\Checkmark &\Checkmark &\XSolidBrush & $96.99$ & $96.83$ & $97.14$ \\
			E&\Checkmark &\Checkmark &\Checkmark & \bm{$98.6$} & \bm{$98.63$} & \bm{$98.57$} \\
			\hline
			\hline
		\end{tabular}
	\end{table}
	
	\begin{figure}[b]
		%\centering	
		%\includegraphics[width=\textwidth]{figures/main_stream.png}
		\includegraphics[width=0.97\columnwidth]{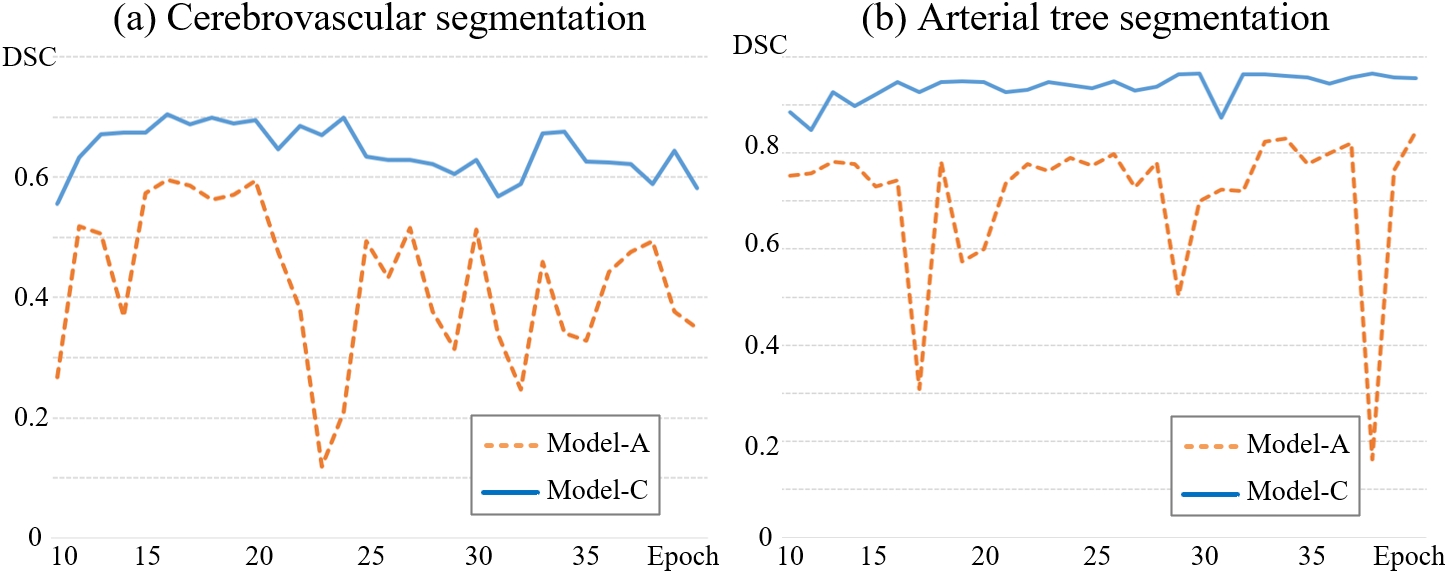}
		\caption{The test DSC for model-A and model-C in epoch 10-40.} 
		\label{fig:ablation_Lcon}
	\end{figure}
	
	\begin{figure}[]
		%\centering	
		%\includegraphics[width=\textwidth]{figures/main_stream.png}
		\includegraphics[width=0.97\columnwidth]{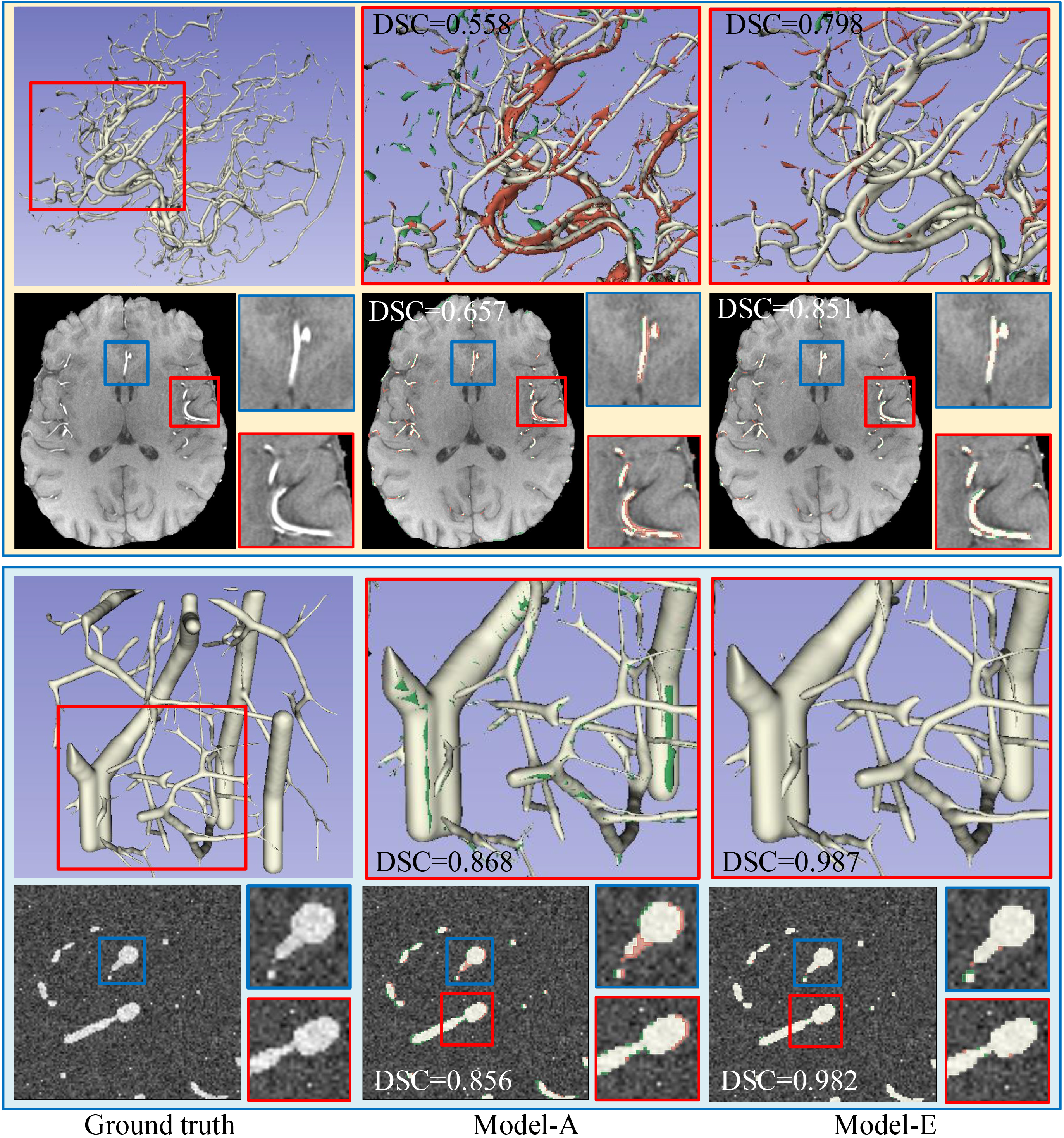}
		\caption{The visualization of cerebrovascular segmentation results and arterial segmentation results produced by model-A and model-E. Regions covered with \textcolor{red}{red} and \textcolor{green}{green} are respectively false negatives and false positives.} 
		\label{fig:ablation_visualization}
	\end{figure}
	
	We respectively analyze the effectiveness of the temporal consistency loss $\mathcal{L}_{con}$, ori-CNN module and Reliability-Refinement algorithm (RR). The results of ablation experiments are shown in Table.~\ref{table:Ablation_brain_vessel}.
	
	\subsubsection{The effectiveness of ori-CNN module} By comparing the results of model-A (baseline) and model-B, we can observe that the proposed ori-CNN module can bring respectively 4.93\% and 10.17\% DSC improvements for the cerebrovascular segmentation and the arterial segmentation. This observation shows that the proposed ori-CNN module, which motivates the model to learn features robust to visual orientation changes, works well in enhancing 3D vascular segmentation models. 
	
	\subsubsection{The effectiveness of $\mathcal{L}_{con}$} In comparison with model-A, model-C trained with $\mathcal{L}_{con}$ can respectively improve the results of cerebrovascular segmentation and arterial segmentation by 14.14\% and 11.0\% in DSC. Meanwhile, as shown by the results of model-D, the implementation of $\mathcal{L}_{con}$ can further improve model-B composed of ori-CNN models. These results show the significant effectiveness of $\mathcal{L}_{con}$. 
	
	To better illustrate the effectiveness of $\mathcal{L}_{con}$ in enhancing the model's training stability, we utilize the historical versions of model-A and model-C in epoch 10-40 to predict one test instance and the test DSC values are shown in Fig.~\ref{fig:ablation_Lcon}. From Fig.~\ref{fig:ablation_Lcon}, the test DSC for models trained without $\mathcal{L}_{con}$ (model-A) fluctuates significantly, which means that the model changes unstably during training. Comparatively, the test DSC for models trained with $\mathcal{L}_{con}$ (model-C) is more stable. This observation demonstrates that $\mathcal{L}_{con}$ can enhance the training stability and ensure the model's smooth update.

	\subsubsection{The effectiveness of Reliability-Refinement algorithm} As shown by the results of model-E and model-D, the implementation of Reliability-Refinement algorithm can further respectively improve model-D by 2.51\% and 1.61\% in the DSC of cerebrovascular segmentation and arterial segmentation. This observation shows the effectiveness of our Reliability-Refinement algorithm and demonstrates that it is effective to utilize the high-confidence predictions as the pseudo labels to further refine the model.
	
	\subsubsection{Analyses on the proposed 3D-SLD model}
	Compared with the baseline (model-A), the 3D-SLD model (model-E) respectively improves the DSC of the cerebrovascular segmentation and arterial segmentation by 22.79\% and 14.16\%, demonstrating the effectiveness of the proposed modules. And according to the results visualization in Fig.~\ref{fig:ablation_visualization}, we can see that the final model can effectively alleviate false negatives and false positives, and works better in the boundary regions and tiny target regions.
	
	\section{Discussion}
	The effectiveness of our 3D-SLD in the 3D vascular segmentation is demonstrated by the experimental validations. Our major findings include: (1) 3D-SLD owns the superior capacity for annotation reuse and can effectively realize segmentation for different 3D vascular structures only under the morphology guidance from public 2D annotations of one vascular structure; (2) 3D-SLD can fully capture the morphology knowledge from 2D vessel masks, which makes it work well with the guidance from annotations of different vascular structures and stably work even with only one reference mask.
	
	Recent deep learning-based 3D vascular segmentation methods rely on the expensive manual annotations specifically designed for the target vascular structure \cite{li20213d,chen2022generative,xia20223d}, which ignores the shared knowledge among different vascular structures and leads to heavy burdens to healthcare systems. The proposed 3D-SLD is the first one to effectively realize 3D vessel segmentation by reusing the public high-quality 2D vessel annotations, which shows essential clinical values to avoid the repetitive annotating and simplify the process of 3D vascular segmentation model building. Firstly, the fusion of the region discrimination loss, adversarial loss and temporal consistency loss contributes to the superior annotation reuse capacity of 3D-SLD and facilitates the stable segmentation of the 3D tree-shaped vessel with internal consistency. Therefore, 3D-SLD can effectively realize the segmentation for different 3D vascular structures, including the 3D cerebral vessels and 3D arterial trees, only based on the annotation of 2D coronary artery masks (shown in Sec.~\ref{sec:experimental-comparisons}). Furthermore, the effectiveness of 3D-SLD is shown by getting results comparable to nine supervised deep learning methods based on associated expensive annotations. Secondly, 3D-SLD is based on the common property that 2D and 3D vessels share the tree-shaped morphology, and the proposed Crop-and-Overlap strategy further constructs reference masks to better represent the morphology knowledge for varied vascular structures. Therefore, 3D-SLD shows the superior capacity of capturing the morphology knowledge, and the 2D annotations of different vascular structures even one 2D structure-agnostic vessel annotation can successfully guide the 3D vascular segmentation (shown in Sec.~\ref{sec:different-reference} and Sec.~\ref{sec:limited-references}).     
	
	\section{Conclusion}
	In this work, we propose the 3D shape-guided local discrimination (3D-SLD) to eliminate the dependence on the expensive and repetitive manual annotations as required by the deep learning-based 3D vascular segmentation models. The 3D-SLD owns the capability for learning discriminative representation from unlabeled images to cluster semantically consistent vascular voxels, capturing the morphology knowledge from 2D vascular annotations to guide the tree-shaped morphology of the segmentation, and alleviating the training unstability by using the temporal information provided by the exponential moving average model. Therefore, the tree-shaped 3D vessels composed of semantically consistent voxels can be stably segmented. Furthermore, the orientation-invariant CNN module and Reliability-Refinement algorithm are presented to successfully enhance the robustness and reliability of the model. The experimental comparisons demonstrate our 3D-SLD guided by public 2D structure-agnostic vessel annotations can effectively obtain results comparable to nine supervised deep learning methods. Moreover, the superior annotation reuse capability enables 3D-SLD to realize segmentation for different 3D structures with the guidance of different 2D reference annotations. Particularly, our model is capable to construct the 3D vessel segmentation model with only one 2D structure-agnostic vessel annotation as the reference.

\end{document}